Markus F. Weber[1,2,#], Gabriele Poxleitner[2,#], Elke Hebisch[2,‡], Erwin Frey[1,2], Madeleine Opitz[2,*]

[1]Arnold Sommerfeld Center for Theoretical Physics, Faculty of Physics, Ludwig-Maximilians-Universität München, Theresienstraße 37, D-80333 Munich, Germany.

[2]Center for NanoScience, Faculty of Physics, Ludwig-Maximilians-Universität München, Geschwister-Scholl-Platz 1, D-80539 Munich, Germany.

[#]These authors contributed equally to this study.

[‡]Current address: Max Planck Institute for Biophysical Chemistry, Department of NanoBiophotonics, Am Fassberg 11, D-37077 Göttingen, Germany.

[*]Opitz@physik.uni-muenchen.de

**Summary**

Dispersal of species is a fundamental ecological process in the evolution and maintenance of biodiversity. Limited control over ecological parameters has hindered progress in understanding of what enables species to colonise new area, as well as the importance of inter-species interactions. Such control is necessary to construct reliable mathematical models of ecosystems. In our work, we studied dispersal in the context of bacterial range expansions and identified the major determinants of species coexistence for a bacterial model system of three *Escherichia coli* strains (toxin producing, sensitive, and resistant). Genetic engineering allowed us to tune strain growth rates and to design different ecological scenarios (cyclic and hierarchical). We found that coexistence of all strains depended on three strongly interdependent factors: composition of inoculum, relative strain growth rates, and effective toxin range. Robust agreement between our experiments and a thoroughly calibrated computational model enabled us to extrapolate these intricate interdependencies in terms of phenomenological biodiversity laws. Our mathematical analysis also suggested that cyclic dominance between strains is not a prerequisite for coexistence in competitive range expansions. Instead, robust three-strain coexistence required a balance between growth rates and either a reduced initial ratio of the toxin-producing strain, or a sufficiently short toxin range.

# 1. Introduction

The fate of a species depends on the abilities of its members to colonise new area and to outperform competitors [1, 2]. A central theme of ecological research is to understand these abilities and to explain how many competing species still manage to live in lasting coexistence, especially during arms races over common resources [3-10]. Structured environments were theoretically proposed to be facilitators of biodiversity [3, 11-19]. However, experimental verification of the proposed mechanisms promoting biodiversity is hard to come by. Ecological studies traditionally focused on systems of mammals and plants, but the long reproduction times and large spatial scales involved impede experimental progress [20]. To circumvent these problems, recent studies have turned to microbial model systems in which both spatial and temporal scales are experimentally better accessible [3, 21-23]. New methods of genetic engineering even admit the possibility to modify the behaviour of test species. These methods stimulated further research on microbial systems and increased our knowledge about their transient and long-term dynamics [24]. For microbial life in well-mixed culture, for example, experimental and theoretical models have recently shown how transient processes can be amplified by recurring life cycles to change a system's long-term fate [25, 26]. In spatial environments, long-term limits are more difficult to attain. We followed a previous study on competitions of three bacterial strains of *Escherichia coli* (toxin producing, sensitive, and resistant) in fixed spatial environments [3] and identified traits that ensure the transient coexistence of strains during the course of range expansions.

What determines whether a bacterial species thrives or falters as it explores new area can be studied systematically after droplet inoculation on an agar plate [5, 27-30]. Recent experimental studies have highlighted the importance of random genetic drift in driving population differentiation along the expanding fronts of bacterial colonies - an effect that gives rise to monoclonal sectoring patterns [5, 31]. Natural microbial colonies and biofilms are characterised by a complex community structure [21, 22, 32], which is shaped by competition between strains for resources such as nutrients and space [2, 5, 27-30], interference competition through the production of toxins [3, 8, 22, 29, 33, 34], and different forms of mutualism, cooperation and cheating [4, 6, 22, 35]. Only a few recent studies, most of them theoretical, have explored the role of such interactions for expanding populations [36-38]. Experimental studies are appearing just recently [10, 39, 40] and are much needed to identify and characterise the key principles that drive population dynamics in expanding systems. In our work, we investigated range expansions for a bacterial model system comprising three *Escherichia coli* strains: a toxic strain, a sensitive strain (facing death upon the encounter of toxins), and a resistant strain. By genetically altering strain growth rates, we created three different ecological scenarios, including a hierarchical scenario and a scenario that mimicked a cyclic rock-paper-scissors game (figure 1) [3, 18]. Control over strain growth rates also enabled us to acquire sufficient experimental data to construct and validate a computational model of the expansion process. The model was used to predict parameter regimes for which coexistence of all three strains was observed in experiments. Furthermore, we identified the factors that determined a strain's chance of survival (composition of the inoculum, relative growth rates, and effective toxin range), and quantified the relationship between these factors in terms of phenomenological "biodiversity laws". Our work highlights the central importance of bacterial interactions in the evolution and maintenance of biological diversity, and pursues the theoretical aim to understand how interactions affect coexistence [41].

## 2. Materials and methods

*(a) Bacterial strains and fluorescent proteins*

The strains used in our study represent the *Escherichia coli* Colicin E2 system (BZB1011 (sensitive "S" strain), $E2^C$-BZB1011 (toxic "C" strain), and $E2^R$-BZB1011 (resistant "R" strain)) [3]. For visualisation of distinct strains, plasmids expressing either the GFP, the red fluorescent protein mCherry (mCh), or no fluorescent protein (nfp) were introduced into S, R and C, respectively. The resulting strains were named: $S_{GFP}$, $S_{mCh}$, $S_{nfp}$, $R_{GFP}$, etc. All fluorescent proteins were expressed from the arabinose inducible promoter pBAD as present in the plasmid pBAD24. Introduction of the fluorescent proteins resulted in the plasmids pBAD24-GFP [42] and pBAD24-mCherry [43]. To prevent plasmid loss, all plasmids, including the plasmid not expressing a fluorescent protein, carried an Ampicillin antibiotic resistance.

*(b) Preparation of the system and growth conditions*

Bacteria were grown in overnight cultures of liquid M63 medium at 37°C, supplemented with glycerol (0.2%), caseinhydrolysat (0.2%), and arabinose (0.2%) for fluorescence induction, and with ampicillin (100 µg/ml). Analysis of colony development was performed on M63 agar plates (1.5% agar) that were prepared as above for the liquid culture.

Strain mixtures were diluted from the overnight culture to OD 0.1 at different initial ratios as indicated in the next sections. Ratios S:R:C (shorthand for $r_S:r_R:r_C$) of 1:1:1 represent an equal amount of all three strains. Ratios 5:1:1 indicate that the S strain was initially added five times more than the R and C strains, while ratios of 1:1:0.1 indicate that the C strain was added at 1/10[th] of the other two strains. Droplets of the resulting mixture (1 µl) were applied to M63 agar plates in triplicate. The time between mixing of strains and inoculation had to be kept short since droplets of inoculum temporarily form well-mixed environments. Tuning the pH level of our agar plates resulted in slow colony growth at pH 6 (slow growth condition "S") compared to fast colony growth at optimal pH 7 (fast growth condition "F"). Each experiment (for slow and fast growth conditions) was performed two times and revealed qualitatively the same result.

*(c) Analysis of colony development*

Colony development was recorded using an upright microscope (90i, Nikon, Düsseldorf, Germany). Fluorescence was analysed using filter sets with 472/30 nm excitation for GFP (DM: 495, BA: 520/35 BP), while excitation for mCherry was 562/25 nm (DM: 593, BA: 641/45 BP). Images were taken with a DS-Qi1MC digital camera (Nikon). Background correction and image analysis were performed using the free software ImageJ.

In order to quantify the growth dynamics of the three strains, we recorded the expansion of single-strain colonies for each combination of strain, fluorescent marker and growth condition in parallel by taking bright field images. For slow growth conditions, these pictures were recorded every two hours from 4 to 34 hours after inoculation. A final picture was recorded 48 hours after inoculation. For fast growth conditions, the pictures were recorded every hour from 4 to 16 hours and every two hours from 16 to 30 hours after inoculation. A final picture was recorded 50 hours after inoculation.

Bright field images were also recorded for the expansion of multi-strain colonies 48 hours after inoculation, together with images for the two fluorescent proteins GFP and mCherry. Only colour overlays of the two fluorescence channels are shown. We chose a natural representation of colours for the visualisation of fluorescent strains.

Strains expressing the green fluorescent protein GFP are, therefore, shown in green, strains expressing the red fluorescent protein mCherry are shown in red. The choice helps to identify the strain that suffered a significant decrease in growth rate due to the expression of mCherry. For slow growth conditions, expression of mCherry by the C strain caused a decrease of its growth rate of 21.5% as compared to its non-fluorescent state, for the R strain a decrease of 22.6%, and for the S strain a decrease of 24.3%. Growth rate decrease was less for fast growth conditions: 13.0% for C, 16.4% for R, and 15.7% for S. All growth rates are listed and displayed in the electronic supplementary material, table S1 and figure S1. A strain not expressing a fluorescent protein (nfp) was present in black areas of a colour overlay. For cases in which the non-fluorescent strain could not be distinguished from surrounding agar, we used a bright field image to delineate the boundary of the colony (cf. figure 2).

*(d) Computational model of competitive range expansions*

Our theoretical model rested on a coarse-grained, mesoscopic description of the bacterial expansion process. The model was agent-based and movement of agents was restricted to a two-dimensional lattice, following previous stochastic simulations of range expansions [31]. Due to the exceedingly large number of bacterial cells in experimentally observed colonies, it was not possible to treat individual bacterial cells as agents. Instead, each agent (a colonised lattice site) represented the bacterial strain that locally dominated a certain area (patch) of a colony (further details on the physical size represented by lattice sites can be found in the electronic supplementary material, appendix S1 and table S1). Our model thereby coarse-grained the microscopic dynamics and reduced both high cell numbers in the lateral direction as well as the increasing number of cell layers in the vertical direction to a lattice of patches.

The growth dynamics of the expansion process was modelled as "hopping processes" from colonised to free patches. The speed of these processes depended on two strain-dependent parameters: mesoscopic growth rate $\mu_m$ and mesoscopic lag-time $\tau_m$. These parameters were adjusted such that our simulation reproduced experimental data on the radial growth of single-strain colonies. The mesoscopic lag-time helped us to effectively add a time dependence to the mesoscopic growth rate (a colonised patch could only proliferate after its lag-time $\tau_m$ had passed). This time dependence allowed us to reproduce also the lag phase and the gradual expansion of colonies during acceleration phase that were observed in growth curves (see electronic supplementary material, figure S8). In order to match simulated growth curves to experimental data, we sampled $\tau_m$ on initially colonised lattice sites from broad, strain and marker dependent Gaussian distributions. The calibration of these distributions is explained in the electronic supplementary material, appendix S1.

The temporal dynamics of range expansions was simulated employing a Gillespie algorithm [44]. The algorithm also governed the toxin interaction between sensitive and toxic lattice sites (we assumed that 3% of initially and newly colonised lattice sites dominated by the C strain were toxic [45]). Our inclusion of this interaction explicitly accounted for the long-range diffusion of colicins (a nearest-neighbour interaction would have been insufficient to recover experimental results). Since diffusion of colicins happens on a much faster time scale than consecutive cell divisions of *E.coli* (diffusion constant of colicins on the order of $10^{-7}$ cm$^2$/s [46]), we approximated the colicin dynamics by a stationary source and degradation process. The process suggested exponentially decaying colicin profiles around toxin producing lattice sites. The colicin profiles introduced two additional parameters into the simulation: their heights around sources (local colicin strength $\kappa$) and their widths (characteristic length scale $\lambda$). We adjusted the parameters using estimates from literature [33, 47] and by calibrating them to experiments on colliding sensitive and colicin producing colonies (see electronic supplementary material, figure S11).

## 3. Results and Discussion

*(a) Design of distinct ecological scenarios*

As detailed in *Materials and methods*, we studied competitive microbial range expansion for a prominent model system that comprises three genetically distinct strains of *Escherichia coli* [3]: a toxin-producing strain (C), a sensitive strain (S), and a resistant strain (R). During growth, around 3% [45] of the C cells undergo lysis and release colicin E2 (diffusion constant on the order of $10^{-7}$ cm$^2$/s [46]). The colicins subsequently diffuse through the extracellular fluid around bacterial cells until possibly being absorbed by sensitive *E. coli* cells. The sensitive cells are prone to the endonuclease activity of colicin and suffer a degeneration of their DNA, which inhibits further cell divisions [48]. Eventually, the cells lyse. Inhibition zones around toxic C cells may be as large as 100-400 µm in radius [33, 47]. Since colicin production is costly, the growth rate of these cells is significantly lower than those of the other two strains. We genetically engineered two of our three strains to express either the green fluorescent protein GFP, or the red fluorescent protein mCherry (the strain not expressing a fluorescent protein is marked as nfp). We observed that while a strain expressing GFP could expand at roughly the same speed as its non-fluorescent counterpart, this did not hold for strains expressing mCherry. We discovered that their growth rate was significantly reduced by the expression of mCherry (see the electronic supplementary material, figure S1*a*). This effect was also observed for growth in liquid culture [43]. The fluorescent proteins thereby allowed us to design different ecological scenarios by changing the order in which the proteins were assigned to our strains (figure 1). Every scenario differed from one another by changes in relative strain growth rates as described in the following. Furthermore, the fluorescent proteins allowed us to visualise each strain independently during its expansion in range (see *Materials and methods*)

The control over the growth rates of our three strains enabled us to design three different ecological scenarios and to study how the composition of expanding colonies depended on the interplay between resource and interference competition. In a first scenario (I), we arranged the bacterial growth rates such that: $\mu_S > \mu_R > \mu_C$ (mCherry expressed by the R strain). As detailed in the electronic supplementary material, appendix S1, we determined these growth rates by measuring the maximal radial expansion velocity (µm/h) of single-strain S, R, and C colonies. These rates were thus independent of the toxin action of C on S, which was quantified independently as described further below. It followed from the above hierarchy $\mu_S > \mu_R > \mu_C$ and from the toxin action of C on S that our first ecological system exhibited a *cyclic* (*non-transitive*) dominance: C dominated S by killing it, S outgrew R, which in turn outgrew C (figure 1*a*). This hierarchy resembled the order of strategies in the children's game rock-paper-scissors [3, 18]. In the second scenario (II), the ordering of growth rates was chosen as: $\mu_R > \mu_S \geq \mu_C$ (mCherry expressed by the S strain). Hence, the interaction network was *strictly hierarchical* (*transitive*), with R displacing C because of its growth advantage, and C displacing S through its allelopathic effect on S (figure 1*b*). In a third intermediate scenario (III), the toxic strain had by far the lowest growth rate, while those of R and S were nearly equal (mCherry expressed by the C strain). Under these conditions the competition network was neither cyclic nor strictly hierarchical: R dominated C by resource competition, and C dominated S by interference competition, but the interaction between R and S was selectively nearly neutral (figure 1*c*). After droplet inoculation of 1 µl mixtures on agar plates (supplemented with minimal M63 medium; see Materials and Methods), we tracked the composition of bacterial colonies over 48 hours and identified the strains that coexisted along expanding fronts of colonies. The strains that were present along these fronts after 48 hours were considered as survivors of the range expansion. It was sufficient that a strain had established at least one stable sector that touched the edge of an expanding colony to be considered a survivor. Our notion of survival and coexistence did not evaluate the number of stable sectors or the relative frequency of strains along the fronts of colonies.

We developed a theoretical model to explain the outcome of bacterial competitions and to predict growth parameters for which a maximal degree of coexistence could be observed in experiments along the expanding fronts of colonies. The predictions were verified experimentally as described in the following. Let us note that we focused on the transient coexistence of bacterial strains on time spans that were accessible to experiments. Korolev et al. developed methods to determine when such transient coexistence is eventually lost [49]. However, the approach does not consider toxin interaction between strains and can only be applied to cases in which either the S or the C strain has already disappeared from a colony's front. In such situations, the strain that eventually dominates may be inferred from the radial expansion velocities of single-strain colonies that are listed in the electronic supplementary material, table S1.

*(b) Cyclic dominance is not sufficient to ensure coexistence of all strains*

We first sought to determine the surviving strains when a droplet of inoculum that contained an equal number of all three strains (initial ratios S:R:C=1:1:1) expanded in range. Surprisingly, in the cyclic rock-paper-scissors scenario I, we found no evidence for coexistence of all three strains. In a previous report such three-strain coexistence was observed for spatially extended systems with a regular arrangement of neighbouring single-strain colonies [3]. Competitive exclusion with dominance of the fastest-growing strain (S) was not observed either. Instead, we found that S was driven to extinction, while strains R and C dominated the colony front, where they formed monoclonal sectors (figure 2*a*). Notably, in the non-cyclic scenarios II and III, coexistence was completely lost. Here, the R strain outcompeted both S and C, and was the only survivor with access to uncolonised area (figures 2*b–c*). Hence, "survival of the fastest" [10] could only be observed in hierarchical scenario II, whereas who survived in the other two scenarios was more subtle and was heavily affected by the long-range toxin action. The outcomes of our bacterial competitions were shown to be robust against small changes in relative growth rates of the strains (induced by reassigning the fluorescent protein GFP while keeping the assignment of mCherry; see the electronic supplementary material, appendix S1), and robust against changes in overall growth conditions (slow growth at pH 6, fast growth at pH 7; see Materials and Methods and the electronic supplementary material, figure S1). The results of supporting experiments are listed in the electronic supplementary material, figures S3 and S4.

*(c) Identification of biodiversity zones*

To elucidate the above findings and to identify the factors that promote or jeopardise survival of the competing strains, we developed a stochastic agent-based model to capture the system dynamics on a coarse-grained scale (see the electronic supplementary material, appendix S1). Our theoretical approach rested on a lattice-based description of range expansions and extended previous models [31] by considering the long-range nature of the toxin interaction. We performed additional experiments on the expansion of single-strain colonies to adjust the model's parameters. Comparisons between experimental and simulated growth curves enabled us to determine all model parameters except for the toxin interaction. We modelled this interaction based on a source and degradation process, and estimated its range and strength by measuring the distance between the front of a growing C colony and the front of a neighbouring S colony (see the electronic supplementary material, appendix S1). The estimate complies with literature values [33, 47]. Even though our theoretical model simplified the bacterial dynamics (e.g. by considering only a single bacterial layer whereas the real colony piled up in hundreds of them in its interior), the model captured the essential parameters. We successfully applied the model to reproduce experimentally observed segregation patterns and to predict the strains that survived a range expansion (figure 2). Let us emphasise that the model's parameterisation rested on independent experiments as described above.

After having established and validated a reliable theoretical model that reproduced our experimental observations, we investigated whether it was possible to rescue the S strain. As the survival of the S strain was directly coupled to the C strain's presence, we analysed how reductions in the initial ratio of the C strain affected the other strains' survival (in particular of the S strain). Simulation data for the cyclic ecological scenario I predicted that reduction of its initial ratio should lead to the formation of broader R sectors at the expense of C (figure 3). The same effect was seen in experiments with initial ratios of S:R:C=1:1:0.5 (see the electronic supplementary material, figure S5). Further reduction of the initial ratio of C in our simulations revealed a regime of three-strain coexistence centred around S:R:C=1:1:0.1 (figure 3). This permissive zone of biodiversity in parameter space coincided remarkably well with our experimental observations of transient three-strain coexistence at ratios 1:1:0.1 (figure 4*a*). For ecological scenarios with a more hierarchical interaction relationship between the strains (scenarios II and III), the R strain was clearly dominant (figures 4*b–c*). Hence, toxin resistance is apparently a more effective survival strategy than either rapid growth or toxin production if the hierarchical order in the competition network is enhanced.

Whether a bacterial strain manages to survive a range expansion and to populate a colony's expanding front depends on two aspects: first, on its ability to form initial clusters in the inoculum from which outward sectors may emerge; second, on the stability of the arising sectors to the annihilation of neighbouring sector boundaries [5]. Both of these aspects are subject to random genetic drift and may prevent the establishment of stable sectors in a simulation (see figure 3). However, whether a specific outcome of the bacterial competition is possible in principle depended solely on the interplay between three factors in our experiments: (i) on the initial strain ratios in the inoculum (demographic noise due to low absolute cell numbers was of minor importance), (ii) on the relative growth rates of the three strains, and (iii) on the effective range of colicin toxicity. On the other hand, differences in lag times between strains played only a minor role in deciding whether a particular strain survived along the expanding front of a colony. To gain insight into the mechanisms responsible for the dependence of biodiversity on the three factors (i)-(iii) and into how they are correlated with each other, we extended our simulations to explore broad parameter ranges.

If the initial ratios of R and C were varied with respect to the initial ratio of S in cyclic ecological scenario I, our simulations showed that biodiversity was most pronounced when the initial ratio $r_C$ of the C strain was reduced to 5-20% of that of the S strain (figure 5*a*). Higher initial ratios of C suppressed growth of the S strain completely, but the R strain ended up dominating the expanding front. In this case, toxin resistance may be seen as a 'cheating' strategy: the R strain could profit from colicin production by the 'cooperating' C strain without having to pay the associated metabolic costs. By cheating, the R strain managed to beat S, even though it would have been the loser in a direct pairwise competition. Conversely, at lower initial ratios of the C strain, the S strain could still bear the incurred costs. Both R and S outgrew the C strain and eventually shared the expanding front. Our results indicated that a narrow range of initial ratios delineated a regime of maximal biodiversity. Biodiversity required that increases in the initial ratio of C were compensated for by even larger increases in the initial ratio of R. The correlation was quantified by the saturation law—a 'biodiversity law'—shown in figure 5*a*. We attributed the saturation to the finite range of colicin toxicity: dense swathes of R cells were needed to shield sensitive cells from the toxin emitted by the C strain. Behind these barriers, surviving S cells could give rise to sectors, leading to the eventual coexistence of all three strains.

Subsequently, we set the initial ratios of the three strains to the rescue window S:R:C=1:1:0.1 and investigated how changes in the relative growth rates of the strains (i.e., changes in the interaction hierarchy) affected the degree of coexistence. Our simulations showed that three-strain coexistence was most pronounced when the growth rates were of comparable size and when the growth rates of strains C and R were varied in a correlated fashion: $\mu_R \sim \mu_C$ (figure 5*b*). As our model predicted two- and three-strain coexistence (as well as its absence) in full accordance with experimental results (R and S in the intermediate scenario III, all three strains in the cyclic scenario I, but only the R strain in the hierarchical scenario II), we expect our theoretical predictions to be highly relevant for future experimental studies. Moreover, our simulations revealed that cyclic dominance is not a necessary prerequisite for biodiversity. For range expansion ecologies, biological diversity can even be maintained if the toxin-producing C strain grows fastest. This result seems paradoxical at first sight, but it demonstrates that both the initial ratios and the growth rates of competing strains are equally important ecological parameters. During the initial phase of expansion after inoculation, the combined effect of the two parameters determines which strain is more likely to establish sector-like domains. In order to avoid being overgrown by the other two strains, the C strain must compensate for its lower initial ratio by growing at a faster rate. A phase diagram that resembled the one in figure 5*b* was computed for range expansions of selectively neutral, non-interacting strains at equal initial strain ratios. The biodiversity window of this null model disappeared in the presence of toxin interaction, but was recovered upon reducing the initial C strain ratio. The changes to the null model were crucial for predicting the surviving strains in our experiments. It would be highly interesting to study how other kinds of bacterial interactions affect the coexistence diagram of the null model.

Finally, to understand the role of colicin in maintaining biodiversity during range expansions, we analysed the importance of the toxin's effective range (see the electronic supplementary material, appendix S1). Our *in silico* studies revealed that maintenance of biodiversity required a strong inverse correlation between the initial ratio of the toxic strain and the length scale of colicin toxicity: $r_C \sim 1 / \lambda^{2.46}$ (figure 5*c*). A long-range toxin interaction (length scale of $\lambda \approx 125$ µm) was, therefore, optimal for species coexistence around the initial strain ratios S:R:C=1:1:0.1. However, our simulations suggested that a more circumscribed radius of toxin action ($\lambda \approx 50$ µm) would be necessary to sustain coexistence at equal strain ratios 1:1:1. The reduction in colicin range weakened the allelopathic effect of C on the fast-growing S strain to a level at which all strains could coexist along the expanding front, despite equal initial ratios in the inoculum. In conclusion, the coexistence diagram in figure 5*c* revealed that changes in the range of colicin toxicity have a strong impact on biodiversity. The maintenance of coexistence relied on the fine-tuning of the interference competition via colicin between strains. In more general terms, the biodiversity law encodes how coexistence depends on the balance between the amount of the producers of an interaction agent and the range of the agent. We expect that the inverse correlation between the two can also be observed in other systems in which an inhibiting interaction is mediated by an agent. Future studies should explore how the law changes for other kind of interactions.

## 4. Conclusion

Range expansion experiments provide a new perspective on the significance of competition between species in spatially extended ecological systems. Neither strength of numbers, nor growth rate differences, nor choice of competition strategy alone determines success of their dispersal. The right balance between these factors must be struck. We identified this balance for range expansions of a bacterial model system of three *Escherichia coli* strains and experimentally validated theoretical predictions on strain coexistence. We used the model to extrapolate in parameter space and described the regimes of maximal coexistence in terms of phenomenological 'biodiversity laws'. The laws showed how changes in the interaction between bacterial strains can have subtle but lasting effects on the eventual composition of a microbial ecosystem. Our approach may help to understand more complex ecosystems whose dynamics cannot be formulated in terms of simplistic rules.

**Acknowledgments**

We thank M. Riley for the kind gift of the original S, R, and C strains, and K. Jung for the kind gift of plasmid pBAD24-GFP and pmCherry. For fruitful discussions and technical support we thank A. Boschini, J.-T. Kuhr, J. Landsberg, G. Schwake, B. Nickel, J. O. Rädler, and T. Reichenbach. The authors also thank D. Braun, J. Knebel, B. Osberg, and P. B. Rainey for discussions and critical reading of the manuscript. This work was supported by the Deutsche Forschungsgemeinschaft through grants FR 850/9-1 and RA 655/5-1, the NanoSystems Initiative Munich (NIM), and the Center for Nanoscience (CeNS).

**Author contributions**

M.O. and E.F. designed the research. G.P. and E.H. performed the experiments. M.F.W. conceived and performed the simulations. G.P., M.F.W., E.F., and M.O. interpreted the data. M.F.W., E.F., and M.O. wrote the paper.

**References**

1. Sakai, A.K., Allendorf, F.W., Holt, J.S., Lodge, D.M., Molofsky, J., With, K.A., Baughman, S., Cabin, R.J., Cohen, J.E., Ellstrand, N.C., et al. 2001 The population biology of invasive species. *Annu. Rev. Ecol. Syst.* **32**, 305-332. (doi:10.1146/Annurev.Ecolsys.32.081501.114037)

2. Hastings, A., Cuddington, K., Davies, K.F., Dugaw, C.J., Elmendorf, S., Freestone, A., Harrison, S., Holland, M., Lambrinos, J., Malvadkar, U., et al. 2005 The spatial spread of invasions: new developments in theory and evidence. *Ecol. Lett.* **8**, 91-101. (doi:10.1111/J.1461-0248.2004.00687.X)

3. Kerr, B., Riley, M.A., Feldman, M.W. & Bohannan, B.J.M. 2002 Local dispersal promotes biodiversity in a real-life game of rock-paper-scissors. *Nature* **418**, 171-174. (doi:10.1038/Nature00823)

4. Greig, D. & Travisano, M. 2004 The prisoner's dilemma and polymorphism in yeast SUC genes. *Proc. R. Soc. B* **271**, S25-S26. (doi:10.1098/Rsbl.2003.0083)

5. Hallatschek, O., Hersen, P., Ramanathan, S. & Nelson, D.R. 2007 Genetic drift at expanding frontiers promotes gene segregation. *Proc. Natl. Acad. Sci. USA* **104**, 19926-19930. (doi:10.1073/Pnas.0710150104)

6. Gore, J., Youk, H. & Van Oudenaarden, A. 2009 Snowdrift game dynamics and facultative cheating in yeast. *Nature* **459**, 253-256. (doi:10.1038/Nature07921)

7. Mougi, A. & Kondoh, M. 2012 Diversity of interaction types and ecological community stability. *Science* **337**, 349-351. (doi:10.1126/Science.1220529)

8. Cordero, O.X., Wildschutte, H., Kirkup, B., Proehl, S., Ngo, L., Hussain, F., Le Roux, F., Mincer, T. & Polz, M.F. 2012 Ecological populations of bacteria act as socially cohesive units of antibiotic production and resistance. *Science* **337**, 1228-1231. (doi:10.1126/Science.1219385)

9. Datta, M.S., Korolev, K.S., Cvijovic, I., Dudley, C. & Gore, J. 2013 Range expansion promotes cooperation in an experimental microbial metapopulation. *Proc. Natl. Acad. Sci. USA* **110**, 7354-7359. (doi:10.1073/pnas.1217517110)

10. Van Dyken, J.D., Müller, M.J.I., Mack, K.M.L. & Desai, M.M. 2013 Spatial Population Expansion Promotes the Evolution of Cooperation in an Experimental Prisoner's Dilemma. *Curr. Biol.* **23**, 919-923. (doi:10.1016/j.cub.2013.04.026)

11. Nowak, M.A. & May, R.M. 1992 Evolutionary games and spatial chaos. *Nature* **359**, 826-829. (doi:10.1038/359826a0)

12. Frank, S.A. 1994 Spatial polymorphism of bacteriocins and other allelopathic traits. *Evol. Ecol.* **8**, 369-386. (doi:10.1007/Bf01238189)

13. Hassell, M.P., Comins, H.N. & May, R.M. 1994 Species coexistence and self-organizing spatial dynamics. *Nature* **370**, 290-292. (doi:10.1038/370290a0)

14. Durrett, R. & Levin, S. 1997 Allelopathy in spatially distributed populations. *J. Theor. Biol.* **185**, 165-171. (doi:10.1006/Jtbi.1996.0292)

15. Czárán, T.L., Hoekstra, R.F. & Pagie, L. 2002 Chemical warfare between microbes promotes biodiversity. *Proc. Natl. Acad. Sci. USA* **99**, 786-790. (doi:10.1073/Pnas.012399899)

16. Amarasekare, P. 2003 Competitive coexistence in spatially structured environments: a synthesis. *Ecol. Lett.* **6**, 1109-1122. (doi:10.1046/J.1461-0248.2003.00530.X)

17. Szabó, G. & Fáth, G. 2007 Evolutionary games on graphs. *Phys. Rep.* **446**, 97-216. (doi:10.1016/J.Physrep.2007.04.004)

18. Reichenbach, T., Mobilia, M. & Frey, E. 2007 Mobility promotes and jeopardizes biodiversity in rock-paper-scissors games. *Nature* **448**, 1046-1049. (doi:10.1038/Nature06095)

19. Wakano, J.Y., Nowak, M.A. & Hauert, C. 2009 Spatial dynamics of ecological public goods. *Proc. Natl. Acad. Sci. USA* **106**, 7910-7914. (doi:10.1073/Pnas.0812644106)

20. Sinervo, B. & Lively, C.M. 1996 The rock-paper-scissors game and the evolution of alternative male strategies. *Nature* **380**, 240-243. (doi:10.1038/380240a0)

21. Jessup, C.M., Kassen, R., Forde, S.E., Kerr, B., Buckling, A., Rainey, P.B. & Bohannan, B.J.M. 2004 Big questions, small worlds: microbial model systems in ecology. *Trends Ecol. Evol.* **19**, 189-197. (doi:10.1016/J.Tree.2004.01.008)

22. Hibbing, M.E., Fuqua, C., Parsek, M.R. & Peterson, S.B. 2010 Bacterial competition: surviving and thriving in the microbial jungle. *Nat. Rev. Microbiol.* **8**, 15-25. (doi:10.1038/Nrmicro2259)

23. Rainey, P.B. & Travisano, M. 1998 Adaptive radiation in a heterogeneous environment. *Nature* **394**, 69-72. (doi:10.1038/27900)

24. Hastings, A. 2004 Transients: the key to long-term ecological understanding? *Trends Ecol. Evol.* **19**, 39-45. (doi:10.1016/J.Tree.2003.09.007)

25. Chuang, J.S., Rivoire, O. & Leibler, S. 2009 Simpson's Paradox in a Synthetic Microbial System. *Science* **323**, 272-275. (doi:10.1126/Science.1166739)

26. Cremer, J., Melbinger, A. & Frey, E. 2012 Growth dynamics and the evolution of cooperation in microbial populations. *Sci. Rep.* **2**, 281. (doi:10.1038/Srep00281)

27. Ben-Jacob, E., Schochet, O., Tenenbaum, A., Cohen, I., Czirok, A. & Vicsek, T. 1994 Generic modeling of cooperative growth-patterns in bacterial colonies. *Nature* **368**, 46-49. (doi:10.1038/368046a0)

28. Golding, I., Cohen, I. & Ben-Jacob, E. 1999 Studies of sector formation in expanding bacterial colonies. *Europhys. Lett.* **48**, 587-593. (doi:10.1209/Epl/I1999-00524-7)

29. Be'er, A., Ariel, G., Kalisman, O., Helman, Y., Sirota-Madi, A., Zhang, H.P., Florin, E.L., Payne, S.M., Ben-Jacob, E. & Swinney, H.L. 2010 Lethal protein produced in response to competition between sibling bacterial colonies. *Proc. Natl. Acad. Sci. USA* **107**, 6258-6263. (doi:10.1073/Pnas.1001062107)

30. Korolev, K.S., Avlund, M., Hallatschek, O. & Nelson, D.R. 2010 Genetic demixing and evolution in linear stepping stone models. *Rev. Mod. Phys.* **82**, 1691-1718. (doi:10.1103/Revmodphys.82.1691)

31. Hallatschek, O. & Nelson, D.R. 2010 Life at the front of an expanding population. *Evolution* **64**, 193-206. (doi:10.1111/J.1558-5646.2009.00809.X)

32. Hansen, S.K., Rainey, P.B., Haagensen, J.A.J. & Molin, S. 2007 Evolution of species interactions in a biofilm community. *Nature* **445**, 533-536. (doi:10.1038/Nature05514)

33. Chao, L. & Levin, B.R. 1981 Structured habitats and the evolution of anticompetitor toxins in bacteria. *Proc. Natl. Acad. Sci. USA* **78**, 6324-6328. (doi:10.1073/Pnas.78.10.6324)

34. Majeed, H., Gillor, O., Kerr, B. & Riley, M.A. 2011 Competitive interactions in Escherichia coli populations: the role of bacteriocins. *ISME J.* **5**, 71-81. (doi:10.1038/Ismej.2010.90)

35. Rainey, P.B. & Rainey, K. 2003 Evolution of cooperation and conflict in experimental bacterial populations. *Nature* **425**, 72-74. (doi:10.1038/Nature01906)

36. Nadell, C.D., Foster, K.R. & Xavier, J.B. 2010 Emergence of spatial structure in cell groups and the evolution of cooperation. *PloS Comput. Biol.* **6**. (doi:10.1371/Journal.Pcbi.1000716)

37. Korolev, K.S. & Nelson, D.R. 2011 Competition and cooperation in one-dimensional stepping-stone models. *Phys. Rev. Lett.* **107**. (doi:10.1103/Physrevlett.107.088103)

38. Kuhr, J.T., Leisner, M. & Frey, E. 2011 Range expansion with mutation and selection: dynamical phase transition in a two-species Eden model. *New J. Phys.* **13**, 113013. (doi:10.1088/1367-2630/13/11/113013)

39. Momeni, B., Brileya, K.A., Fields, M.W. & Shou, W. 2013 Strong inter-population cooperation leads to partner intermixing in microbial communities. *eLife*, 2:e00230. (doi:10.7554/eLife.00230)

40. Müller, M.J.I., Neugeboren, B.I., Nelson, D.R. & Murray, A.W. 2014 Genetic drift opposes mutualism during spatial population expansion. *Proc. Natl. Acad. Sci. USA* **111**, 1037-1042. (doi:10.1073/Pnas.1313285111)

41. Knebel, J., Krüger, T., Weber, M.F. & Frey, E. 2013 Coexistence and Survival in Conservative Lotka-Volterra Networks. *Phys. Rev. Lett.* **110**. (doi:10.1103/Physrevlett.110.168106)

42. Megerle, J.A., Fritz, G., Gerland, U., Jung, K. & Rädler, J.O. 2008 Timing and dynamics of single cell gene expression in the arabinose utilization system. *Biophys. J.* **95**, 2103-2115. (doi:10.1529/Biophysj.107.127191)

43. Hebisch, E., Knebel, J., Landsberg, J., Frey, E. & Leisner, M. 2013 High Variation of Fluorescence Protein Maturation Times in Closely Related Escherichia coli Strains. *PLoS ONE* **8**, e75991. (doi:10.1371/journal.pone.0075991)

44. Gillespie, D.T. 1976 A general method for numerically simulating the stochastic time evolution of coupled chemical reactions. *J. Comput. Phys.* **22**, 403-434. (doi:10.1016/0021-9991(76)90041-3)

45. Cascales, E., Buchanan, S.K., Duche, D., Kleanthous, C., Lloubes, R., Postle, K., Riley, M., Slatin, S. & Cavard, D. 2007 Colicin biology. *Microbiol. Mol. Biol. R.* **71**, 158-229. (doi:10.1128/Mmbr.00036-06)

46. Liang, S.M., Xu, J., Weng, L.H., Dai, H.J., Zhang, X.L. & Zhang, L.N. 2006 Protein diffusion in agarose hydrogel in situ measured by improved refractive index method. *J. Control. Release* **115**, 189-196. (doi:10.1016/J.Jconrel.2006.08.006)

47. Ozeki, H., Stocker, B.A.D. & De Margerie, H. 1959 Production of colicine by single bacteria. *Nature* **184**, 337-339. (doi:10.1038/184337a0)

48. Schaller, K. & Nomura, M. 1976 Colicin-E2 Is a DNA Endonuclease. *Proc. Natl. Acad. Sci. USA* **73**, 3989-3993. (doi:10.1073/Pnas.73.11.3989)

49. Korolev, K.S., Müller, M.J.I., Karahan, N., Murray, A.W., Hallatschek, O. & Nelson, D.R. 2012 Selective sweeps in growing microbial colonies. *Phys. Biol.* **9**, 026008. (doi:10.1088/1478-3975/9/2/026008)

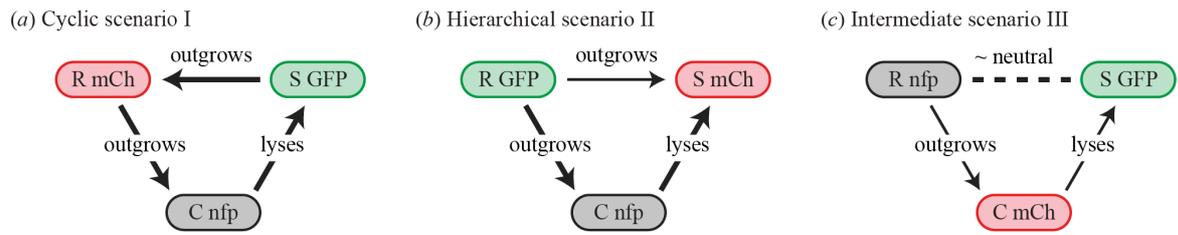

**Figure 1.** Three ecological scenarios that were designed by altering strain growth rates through the expression of the fluorescent protein mCherry. (*a*) In the cyclic scenario I, the hierarchy of single strain growth rates was $\mu_S > \mu_R > \mu_C$ (for numerical values see the electronic supplementary material, table S1 and figure S1). Cyclic dominance held because colicin emitted by the toxin producing C strain inhibited, and eventually lysed, cells of the sensitive S strain. (*b*) In the hierarchical scenario II, the resistant strain outperformed the two other strains and the growth rate hierarchy was $\mu_R > \mu_S \geq \mu_C$. (*c*) In the intermediate scenario III, colonies formed by either the S or the R strain expanded at roughly the same rate, and outgrew colonies formed by the toxin-producing C strain.

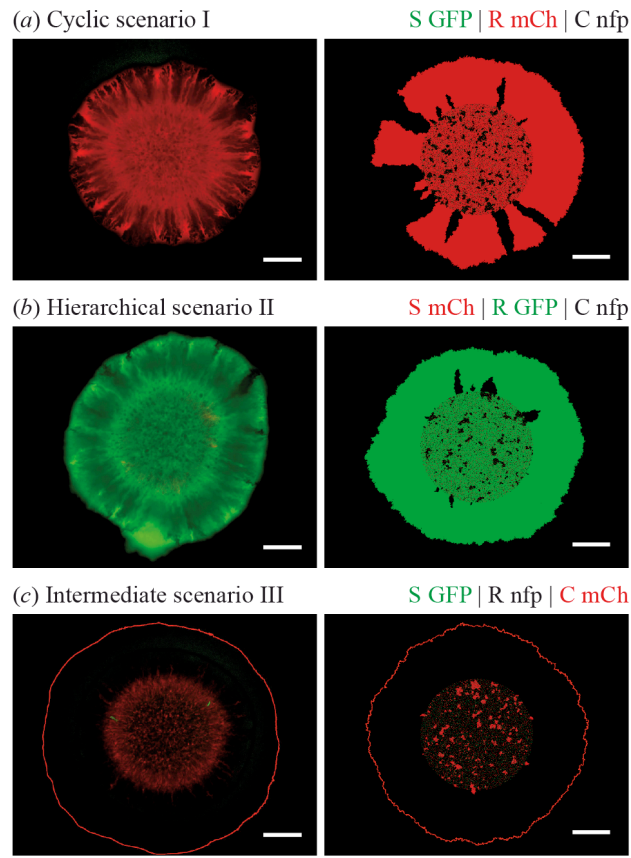

**Figure 2.** Segregation patterns arising in range expansions initiated at initial strain ratios S:R:C=1:1:1. Experimental outcomes are displayed in the left column (images obtained 48 h after inoculation), simulation results in the right column (simulations stopped after colony had reached a radius of 3 mm). Different strain/fluorescent marker combinations were used for visualisation and to implement distinct ecological scenarios. The combinations are indicated above individual rows (GFP: green fluorescent protein, mCh: red fluorescent protein mCherry, nfp: no fluorescent protein). For further information on the robustness of experimental as well as theoretical results see the electronic supplementary material, appendix S1 and figures S3 and S4. White scale bars represent 1 mm. (*a*) Transient coexistence of the R and the C strain was maintained in the cyclic scenario I. (*b–c*) The R strain outcompeted both the S and the C strain in the strictly hierarchical scenario II and in the intermediate scenario III.

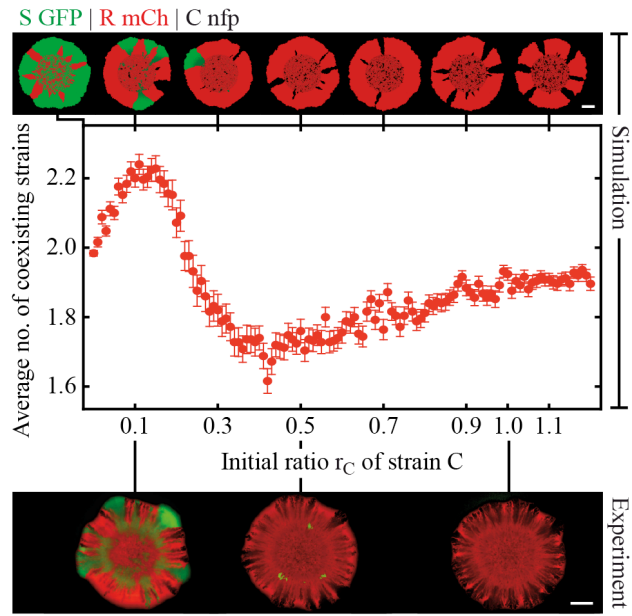

**Figure 3.** Dependence of the degree of coexistence on the relative initial amount of the C strain. With initial ratios of S:R:C=1:1:1 ($r_S$:$r_R$:$r_C$) in the cyclic scenario I, the R and C strain came to dominate the expanding front. As the initial ratio of the C strain $r_C$ was reduced, R strain sectors became broader, in agreement with experimental observations (depicted below the theoretical results; see also the electronic supplementary material, figure S5). Further reduction of the initial ratio of C weakened its allelopathic effect on S such that expanding S sectors emerged. A maximal number of coexisting strains along colony fronts was observed around S:R:C=1:1:0.1 (24% of simulations exhibited three-strain coexistence; error bars: s.e.m., n=250). The three-strain coexistence at this initial ratio was also observed in experiments. White scale bars represent 1 mm.

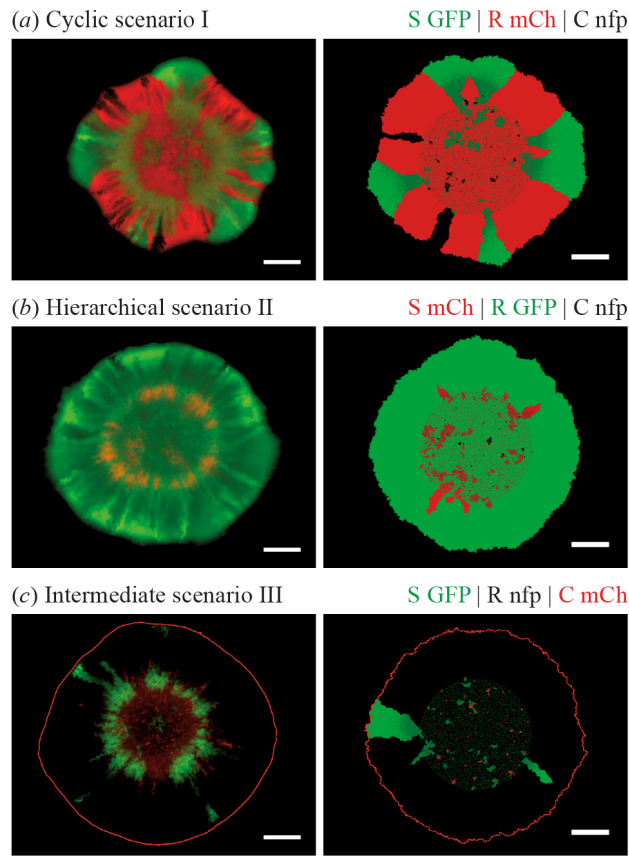

**Figure 4.** Range expansions at initial ratios S:R:C=1:1:0.1. See the legend of figure 2 for additional information. (*a*) A transient coexistence of all three strains was observed along the colony's expanding rim in the cyclic scenario I. The distinct protrusions formed by $S_{GFP}$ hint at its selective advantage over the other two strains (see the electronic supplementary material, table S1). Simulations and deterministic analysis anticipated an eventual dominance of $S_{GFP}$ on longer time-scales [49]. (*b*) In the strictly hierarchical scenario II, the growth rate of the S strain was slowed by the expression of mCherry such that it could not compete against R, despite the low initial ratio of C. (*c*) Both R and S strains survived the range expansion in the intermediate scenario III, with the former strain being dominant over the latter.

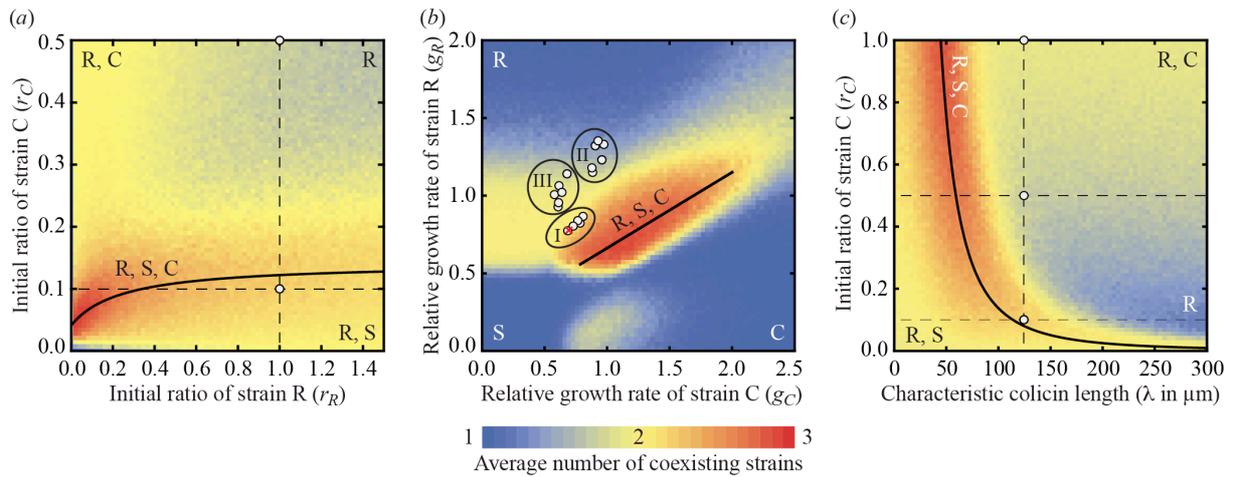

**Figure 5.** Coexistence diagrams and formulation of 'biodiversity laws'. We extended the simulations to study the individual impact of initial strain ratios, relative growth rates, and toxin range. Parameters not under consideration were set to their values in the cyclic scenario I (see the electronic supplementary material, table S1). Letters within the figures indicate strains that survived a range expansion in the corresponding parameter regime. The colour scale on the bottom visualises the level of coexistence averaged over 250 simulations. Solid lines delineate regimes of maximal three-strain coexistence and are referred to as *'biodiversity laws'*. (*a*) Studying the initial ratios of strain C ($r_C$) and of strain R ($r_R$) revealed a regime of maximal coexistence that followed the *saturation law*: $r_C = (0.01 + 0.14\, r_R) / (0.24 + r_R)$. Three-strain coexistence at the lower white circle was supported by experimental realisations. At the upper white circle, survival of R and C was seen in experiments. (*b*) Varying the relative growth rates of strains R and C with respect to the one of S revealed that cyclic dominance is not a prerequisite for the maintenance of biodiversity (initial ratios S:R:C=1:1:0.1). Maximal coexistence follows the *linear law* $g_R = 0.17 + 0.49\, g_C$ (straight line). White circles represent experimental results that were in accordance with our theoretical predictions for the indicated ecological scenario (including experimental results for fast growth conditions and for small

growth rate variations; see the electronic supplementary material, appendix S1). The red cross indicates an experiment in which stable coexistence was not observed ($F_{0.1}$ in the electronic supplementary material, figure S7). Surviving strains were also required to have expanded by at least half the distance of the leading strain out of an inoculum. The criterion was only relevant for highly diverging growth rates (e.g., in the lower left). (*c*) Coexistence diagram for the influence of the C strain's toxicity. The colicin interaction with characteristic length scale of $\lambda$ = 125 μm led to survival of both R and C at initial ratios S:R:C=1:1:1 and 1:1:0.5, and to three-strain coexistence at 1:1:0.1 in experiments (white circles). Simulations predicted an increased level of coexistence along the *power law* $r_C \sim 1 / \lambda^{2.46}$.